\documentstyle[12pt]{article}
\newfont{\ffont}{msym10}                        
\newcommand{\beq}{\begin{equation}}             
\newcommand{\eeq}{\end{equation}}               
\newcommand{\bqry}{\begin{eqnarray}}            
\newcommand{\eqry}{\end{eqnarray}}              
\newcommand{\bqryn}{\begin{eqnarray*}}          
\newcommand{\eqryn}{\end{eqnarray*}}            
\newcommand{\NL}{\nonumber \\}                  
\newcommand{\preprint}[1]{\begin{table}[t]      
            \begin{flushright}                  
            \begin{large}{#1}\end{large}        
            \end{flushright}                    
            \end{table}}                        
\newcommand{\PD}[2]                             
    {\frac{\partial^{#2}}{\partial #1^{#2}}}    
\begin{document}
\preprint{TAUP-2136-94 \\ }
\title{Galilean Limit of
Equilibrium Relativistic Mass Distribution for Indistinguishable Events}
\author{\\ L. Burakovsky\thanks {Bitnet: BURAKOV@TAUNIVM.TAU.AC.IL.} \
and L.P. Horwitz\thanks
  {Bitnet: HORWITZ@TAUNIVM.TAU.AC.IL. Also at Department of Physics,
  Bar-Ilan
University, Ramat-Gan, Israel  } \\ \ }
\date{School of Physics and Astronomy \\ Raymond and Beverly Sackler
Faculty of Exact Sciences \\ Tel-Aviv University,
Tel-Aviv 69978, ISRAEL}
\maketitle
\begin{abstract}
The relativistic distribution for indistinguishable events is considered
in the mass-shell limit $m^2\cong M^2,$ where $M$ is a given intrinsic
property of the events. The characteristic thermodynamic quantities are
calculated and subject to the zero-mass and the high-temperature limits.
The results are shown to be in agreement with the corresponding
expressions of an on-mass-shell relativistic kinetic theory. The Galilean
limit $c\rightarrow \infty ,$ which coincides in form with the
low-temperature limit, is considered. The theory is shown to pass over to
a nonrelativistic statistical mechanics of indistinguishable particles.
\end{abstract}


\newpage
\section{Introduction}
In this paper we consider the mass-shell and the Galilean limits of
equilibrium relativistic distribution for
indistinguishable events studied in a previous work \cite{BH1}. In
that work we studied an identical many-body system within the framework
of a manifestly covariant relativistic statistical mechanics discussed in
a series of papers [2]-[4]. In this framework, for an $N$-body system,
the $N$ $events$ generating the $N$ particle world lines are considered
as the fundamental dynamical objects of the theory; they are
characterized by positions $q^\mu =(ct,{\bf q})$ and energy-momenta $p^
\mu =(E/c,{\bf p})$ in an $8N$-dimensional phase space. Their motion is
parametrized by a continuous Poincar\'{e}-invariant parameter called the
historical time. Such a system of $N$ events is described by generalized
Boltzmann equation [5], whose equilibrium solution gives the distribution
functions (for both bosonic and fermionic events treated simultaneously
in one expression) coinciding with the corresponding grand canonical
distributions obtained in ref. [2] for the static Gibbs ensembles. Upon
integration of this distribution function over angular and hyperangular
variables, one obtains the relativistic mass distribution [1]. We found
expressions for the pressure and the energy density in such a system and
obtained the relativistic equation of state. Now we turn to a mass-shell
form of that equilibrium distribution and to its Galilean
(nonrelativistic) limit.

The Galilean limit of a manifestly covariant relativistic statistical
mechanics was considered in refs. \cite{HR},\cite{BH3}, by taking
$c\rightarrow \infty$ (compared to
all other velocities). In this limit the relativistic relation
between the energy $E$ and the mass $m$ $$E^2=m^2+{\bf p}^2$$
transforms to
\beq
E=m+\frac{{\bf p}^2}{2m}.
\eeq
If we require in addition [6],[7], that the quantity
\beq
\eta =c^2(m-M)
\eeq
may take any value, however, finite, as $c
\rightarrow \infty ,$ then $m=M(1+O(1/c^2))$ \\ (i.e., the ``mass-shell
limit''), and the relation between $E$ and $m$ takes on the form [7]
\beq
E=m+\frac{{\bf p}^2}{2M},
\eeq
where the Galilean mass $M$ coincides with the particle's intrinsic
parameter.

In case of an equilibrium relativistic ensemble of distinguishable events
such a transformation of the relativistic relation between $E$ and $m$
gives rise to the usual nonrelativistic Maxwell-Boltzmann distribution
of ${\bf p}^2/2M$ [7].

In this work, we show that in the mass-shell limit, the equilibrium
relativistic distribution for indistinguishable events used in ref. [1]
approaches the distribution found by J\"{u}ttner and Synge \cite{GLW}
within the framework of an on-mass-shell relativistic kinetic theory. We
also study the $c\rightarrow \infty $ limit of the on-mass-shell theory,
and show that it goes smoothly to the Galilean form.

\section{Equilibrium relativistic distribution close to mass shell}
Consider the equilibrium
relativistic distribution for indistinguishable events used in ref. [1]
(we use the metric $g^{\mu \nu }=(-,+,+,+)$ and $q\equiv q^\mu ,\;p\equiv
p^\mu ,$ and assume no degeneracy; in the case of degeneracy all the
corresponding formulas throughout the paper should be multiplied by
degeneracy factor),
\beq
f_0(q,p)=C(q)\frac{e^{A(q)(p+p_c)^2+B(q)}}{1\mp e^{A(q)(p+p_c)^2+B(q)}},
\;\;\;A(q)>0,
\eeq
which is normalized as
\beq
\int d^4pf_0(q,p)=n(q),
\eeq
where $n(q)$ is the total number of events
per unit space-time volume in the neighborhood of the point $q.$

It follows from the relations [7] (we suppress $c$ for the present
consideration) $$\eta =m-M,\;\;\;
-\triangle \leq \eta \leq \triangle $$ that
\beq
M-\triangle \leq m\leq M+\triangle .
\eeq
Since $\triangle $ may take any value as small as one wishes,\footnote{It
corresponds to the approximate mass shell condition $\mid m-M\mid \leq
\triangle .$} but not zero, i.e., the variation in mass of the particles
of the ensemble may be very small, we can take the value of $p^2\equiv -
m^2$ restricted to a small neighborhood of a fixed value $-M^2.$
This permits us to write (4) as \cite{Jut}
\beq
f_0(q,p)\cong C(q)\frac{e^{-A(M^2+m_c^2)+B}e^{2Ap^\mu p_{c\mu }}}{1\mp e^
{-A(M^2+m_c^2)+B}e^{2Ap^\mu p_{c\mu }}}.
\eeq
Introducing hyperbolic variables [3] and performing integration
\cite{GrRy}, we obtain from (5) and (7) the normalization relation
\beq
n(q)=C(q)\frac{4\pi \triangle M^2}{Am_c}\sum _{n=1}^\infty \frac{(\pm 1)^
{n+1}}{n}e^{-nA(M^2+m_c^2)+nB}K_1(2nAMm_c),
\eeq
where $K_1$ is a Bessel function of the third kind, defined by
\beq
K_{\nu }(z)=\frac{\pi i}{2}e^{\pi i\nu /2}H_{\nu }^{(1)}(iz).
\eeq

The average values of $p^\mu ,\;p^\mu p^\nu ,$ etc., can be obtained from
the corresponding relations of ref. [1], using the formula \cite{GrRy1}
\beq
K_n^\prime (z)=-K_{n+1}(z)+\frac{n}{z}K_n(z).
\eeq
Thus
\beq
\langle p^\mu \rangle _q=p_c^\mu \frac{M}{m_c}\frac{\sum _{n=1}^\infty \{
(\pm 1)^{n+1}e^{-nA(M^2+m_c^2)+nB}K_2(2nAMm_c)/n\}}{\sum _{n=1}^\infty \{
(\pm 1)^{n+1}e^{-nA(M^2+m_c^2)+nB}K_1(2nAMm_c)/n\}},
\eeq
\bqry
\langle p^\mu p^\nu \rangle _q & = & g^{\mu \nu }\frac{M}{2Am_c}\frac
{\sum _{n=1}^\infty \{(\pm 1)^{n+1}e^{-nA(M^2+m_c^2)+nB}K_2(2nAMm_c)/n^2
\}}{\sum _{n=1}^\infty \{(\pm 1)^{n+1}e^{-nA(M^2+m_c^2)+nB}K_1(2nAMm_c)/n
\}} \NL   &   & +M^2\frac{p_c^\mu p_c^\nu }{m_c^2}\frac{\sum _{n=1}^
\infty \{(\pm 1)^{n+1}e^{-nA(M^2+m_c^2)+nB}K_3(2nAMm_c)/n\}}{\sum _{n=1}^
\infty \{(\pm 1)^{n+1}e^{-nA(M^2+m_c^2)+nB}K_1(2nAMm_c)/n\}}.
\eqry
Identifying, as in [1]\footnote{The first relation implies that in
thermal equilibrium $Am_c$ is independent of $q.$},
\bqry
2Am_c & = & \frac{1}{k_BT}, \NL
m_c & = & \frac{M}{\mu _K}, \NL
B & = & \frac{1}{k_BT}\left(\mu +\frac{m_c}{2}\right),
\eqry
where $T,$ $\mu $ and $\mu _K$ are absolute temperature, chemical and
mass potentials, respectively, we have
\beq
\langle p^\mu \rangle =p_c^\mu \frac{M}{m_c}\frac{\sum _{n=1}^\infty \{
(\pm 1)^{n+1}e^{\frac{n\mu ^{'}}{k_BT}}K_2(\frac{nM}{k_BT
})/n\}}{\sum _{n=1}^\infty \{(\pm 1)^{n+1}e^{\frac{n\mu ^{'}}{k_BT}}
K_1(\frac{nM}{k_BT})/n\}},
\eeq
\bqry
\langle p^\mu p^\nu \rangle  & = & g^{\mu \nu }Mk_BT\frac{\sum _{n=1}^
\infty \{(\pm 1)^{n+1}e^{\frac{n\mu ^{'}}{k_BT}}K_2(\frac{nM}{k_BT})
/n^2\}}{\sum _{n=1}^\infty \{(\pm 1)^{n+1}e^{\frac{n\mu ^{'}}{k_BT}}K_1(
\frac{nM}{k_BT})/n\}} \NL   &   & +M^2\frac{p_c^\mu p_c^\nu }{m_c^2}
\frac{\sum _{n=1}^\infty \{(\pm 1)^{n+1}e^{\frac{n\mu ^{'}}{k_BT}}K_3(
\frac{nM}{k_BT})/n\}}{\sum _{n=1}^\infty \{(\pm 1)^{n+1}e^{\frac{n\mu ^
{'}}{k_BT}}K_1(\frac{nM}{k_BT})/n\}},
\eqry
where $\mu ^{'}\equiv \mu -\frac{\mu _KM}{2N}$ $(N$ being the total
number of events) is the ``reduced'' chemical potential approaching $\mu
$ as $N\rightarrow \infty $ or $T\rightarrow 0$ (since in the latter case
$\mu _K\rightarrow 0$ [1]).

As in previous works [1],[3],[4], to
obtain the local energy density we make a Lorentz transformation to
the rest frame of the local average motion. According
to (14), the relative velocity of the new frame is $${\bf u}=\frac{{\bf p
_c}}{m_c}.$$ The rest frame energy is then $$\langle E^{'}\rangle =\frac
{\langle E\rangle -{\bf u}\cdot {\bf p}}{\sqrt{1-{\bf u}^2}},$$ so that
\beq
\langle E^{'}\rangle =M\frac{\sum _{n=1}^\infty \{(\pm 1)^{n+1}e^{\frac{n
\mu ^{'}}{k_BT}}K_2(\frac{nM}{k_BT})/n\}}{\sum _{n=1}^\infty
\{(\pm 1)^{n+1}e^{\frac{n\mu ^{'}}{k_BT}}K_1(\frac{nM}{k_BT})/n\}}.
\eeq

To obtain the pressure and the energy density in our ensemble,
as in previous works [1],[3],[4], we study the $particle$ energy-momentum
tensor defined by the $R^4$ density,
\beq
T^{\mu \nu }(q)=\sum _{i}\int d\tau \frac {p^{\mu }_i p^{\nu }_i }{M}
\delta ^4 (q-q_i(\tau )).
\eeq
Using the result of [3]
\beq
\langle T^{\mu \nu }(q)\rangle _q =T_{\triangle V}\int d^4 pf_0 (q,p)
\frac {p^\mu p^\nu }{M}
\eeq
and the expression (15) for $\langle p^\mu p^\nu \rangle _q $, we obtain
\bqry
\langle T^{\mu \nu }(q)\rangle _q  & = & \frac {T_{\triangle V}n(q)}{M}
\left[g^{\mu \nu }\frac{M}{2Am_c}\frac
{\sum _{n=1}^\infty \{(\pm 1)^{n+1}e^{-nA(M^2+m_c^2)+nB}K_2(2nAMm_c)/n^2
\}}{\sum _{n=1}^\infty \{(\pm 1)^{n+1}e^{-nA(M^2+m_c^2)+nB}K_1(2nAMm_c)/n
\}}\right. \NL  &  & \left. +M^2\frac{p_c^\mu p_c^\nu }{m_c^2}\frac{\sum
_{n=1}^
\infty \{(\pm 1)^{n+1}e^{-nA(M^2+m_c^2)+nB}K_3(2nAMm_c)/n\}}{\sum _{n=1}^
\infty \{(\pm 1)^{n+1}e^{-nA(M^2+m_c^2)+nB}K_1(2nAMm_c)/n\}}\right].
\eqry
In this expression $T_{\triangle V}$ is the average passage interval in
$\tau $ for the events which pass through a small (typical) four-volume
$\triangle V $ in the neighborhood of the point $q$ of $R^4$.

The formula for the stress-energy tensor of a perfect fluid has the form
[3]
\beq
\langle T^{\mu \nu }(q)\rangle _q =pg^{\mu \nu }-(p+\rho )\frac {\langle
 p^{\mu }\rangle _q \langle p^{\nu }\rangle _q }
{\langle p^{\lambda }\rangle _q \langle p_{\lambda }\rangle _q},
\eeq
where $p$ is the pressure and $\rho $ is the density of energy at $q$.

According to  (14),
\beq
\frac{\langle p^{\mu }\rangle _q }
{\sqrt{-\langle p^{\lambda }\rangle _q \langle p_{\lambda }\rangle _q }}
=\frac{p^{\mu }_c}{m_c},
\eeq
hence
\beq
p=\frac{T_{\triangle V}n(q)}{2Am_c}\frac{\sum _{n=1}^\infty \{(\pm 1)^
{n+1}e^{-nA(M^2+m_c^2)+nB}K_2(2nAMm_c)/n^2\}}{\sum _{n=1}^\infty
\{(\pm 1)^{n+1}e^{-nA(M^2+m_c^2)+nB}K_1(2nAMm_c)/n\}}
\eeq
and
\beq
p+\rho =T_{\triangle V}n(q)M\frac{\sum _{n=1}^\infty
\{(\pm 1)^{n+1}e^{-nA(M^2+m_c^2)+nB}K_3(2nAMm_c)/n\}}{\sum _{n=1}^
\infty \{(\pm 1)^{n+1}e^{-nA(M^2+m_c^2)+nB}K_1(2nAMm_c)/n\}}.
\eeq
To interpret these results, as in [1],[3],[4], we calculate the average
(conserved) $particle$ four-current having the microscopic form
\beq
J^{\mu }(q)=\sum _i \int \frac {p^{\mu }_i}{M}\delta ^4 (q-q_i(\tau ))d
\tau .
\eeq
Using the result of [3]
\beq
\langle J^{\mu }(q)\rangle _q =T_{\triangle V}\int d^4 p\frac{p^{\mu }}
{M}f_0 (q,p)
\eeq
and expression (14) for $\langle p^{\mu }\rangle _q$, we obtain
\beq
\langle J^\mu (q)\rangle _q=T_{\triangle V}n(q)\frac{p^{\mu }_c}{m_c}
\frac{\sum _{n=1}^\infty
\{(\pm 1)^{n+1}e^{-nA(M^2+m_c^2)+nB}K_2(2nAMm_c)/n\}}{\sum _{n=1}^\infty
\{(\pm 1)^{n+1}e^{-nA(M^2+m_c^2)+nB}K_1(2nAMm_c)/n\}}.
\eeq
In the local rest frame $p^{\mu }_c =(m_c,{\bf 0})$,
\beq
\langle J^0(q)\rangle _q=T_{\triangle V}n(q)\frac{\sum _{n=1}^\infty
\{(\pm 1)^{n+1}e^{-nA(M^2+m_c^2)+nB}K_2(2nAMm_c)/n\}}{\sum _{n=1}^\infty
\{(\pm 1)^{n+1}e^{-nA(M^2+m_c^2)+nB}K_1(2nAMm_c)/n\}}.
\eeq
Defining the density of $particles$ per unit space volume as
\beq
N_{0}(q)=\langle J^{0}(q)\rangle _q,
\eeq
we obtain the relativistic equation of state
\bqry
p & = & \frac{N_0}{2Am_c}\frac{\sum _{n=1}^\infty \{(\pm 1)^{n+1}
e^{-nA(M^2+m_c^2)+nB}K_2(2nAMm_c)/n^2\}}{\sum _{n=1}^\infty
\{(\pm 1)^{n+1}e^{-nA(M^2+m_c^2)+nB}K_2(2nAMm_c)/n\}} \NL
& = & N_0k_BT\frac{\sum _{n=1}^\infty \{(\pm 1)^{n+1}e^{\frac{n\mu ^{'}}
{k_BT}}K_2(\frac{nM}{k_BT})/n^2\}}{\sum _{n=1}^\infty
\{(\pm 1)^{n+1}e^{\frac{n\mu ^{'}}{k_BT}}K_2(\frac{nM}{k_BT})/n\}}.
\eqry
It follows from (22),(23) that
\beq
\frac{\rho }{p}=\frac{M}{k_BT}\frac{\sum _{n=1}^\infty \{(\pm 1)^{n+1}e^{
\frac {n\mu ^{'}}{k_BT}}K_3(\frac{nM}{k_BT})/n\}}{\sum _{n=1}^\infty \{
(\pm 1)^{n+1}e^{\frac{n\mu ^{'}}{k_BT}}K_2(\frac{nM}{k_BT})/n^2\}}-1.
\eeq
Let us introduce $\Gamma $ through the relation\footnote{This relation is
helpful when studying an adiabatic equation of state \cite{BH4}.}
\beq
p=(\Gamma -1)\rho .
\eeq
Then
\beq
\Gamma =1+\frac{1}{\frac{M}{k_BT}\frac{\sum _{n=1}^\infty \{(\pm 1)^{n+1}
e^{\frac{n\mu ^{'}}{k_BT}}K_3(\frac{nM}{k_BT})/n\}}{\sum _{n=1}^\infty \{
(\pm 1)^{n+1}e^{\frac{n\mu ^{'}}{k_BT}}K_2(\frac{nM}{k_BT})/n^2\}}-1}.
\eeq
Using the asymptotic formulas \cite{AS}
\beq
K_\nu (z)\sim \left\{ \begin{array}{ll}
\sqrt{\frac{\pi }{2z}}e^{-z}, & z\rightarrow \infty , \\
\frac{1}{2}\Gamma (\nu )\left(\frac{z}{2}\right)^{-\nu }\!\!, &
z\rightarrow 0,
\end{array} \right.
\eeq
one obtains
\beq
\Gamma =\left\{ \begin{array}{ll}
1, & T\rightarrow \;0\;, \\
\frac{4}{3}, & T\rightarrow \infty .
\end{array} \right.
\eeq
Since in thermal equilibrium there is no dependence on $q,$ instead of $f
_0(q,p)$ one can use equilibrium relativistic distribution function [5]
\beq
f_0(p)=\frac{1}{e^{(E-\mu +\mu _K\frac{m^2}{2M})/k_BT}\mp 1},
\eeq
normalized as\footnote{Eq. (36) is formally equivalent
to (4),(5), with $C(q)=\frac{1}{(2\pi )^4}.$}
\beq
\int \frac{d^4p}{(2\pi )^4}f_0(p)=n,
\eeq
where $n\equiv \frac{N}{V^{(4)}}$ is the event number density. In this
case one obtains, instead of (8),
\beq
n=\frac{\triangle M^2k_BT}{2\pi ^3}\sum _{n=1}^\infty \frac{(\pm 1)^{n+1}
}{n}e^{\frac{n\mu ^{'}}{k_BT}}K_1\left(\frac{nM}{k_BT}\right),
\eeq
and then, from (22),(23),(27),
\beq
p=\frac{(T_{\triangle V}\triangle )M^2(k_BT)^2}{2\pi ^3}\sum _{n=1}^
\infty \frac{(\pm 1)^{n+1}
}{n^2}e^{\frac{n\mu ^{'}}{k_BT}}K_2\left(\frac{nM}{k_BT}\right),
\eeq
\beq
p+\rho =\frac{(T_{\triangle V}\triangle )M^3k_BT}{2\pi ^3}\sum _{n=1}
^\infty \frac{(\pm 1)^{n+1}
}{n}e^{\frac{n\mu ^{'}}{k_BT}}K_3\left(\frac{nM}{k_BT}\right),
\eeq
\beq
N_0=\frac{(T_{\triangle V}\triangle )M^2k_BT}{2\pi ^3}\sum _{n=1}
^\infty \frac{(\pm 1)^{n+1}
}{n}e^{\frac{n\mu ^{'}}{k_BT}}K_2\left(\frac{nM}{k_BT}\right).
\eeq
Using now the relation \cite{GrRy3}
\beq
K_{n+1}(z)-K_{n-1}(z)=\frac{2n}{z}K_n(z),
\eeq
one has from (38),(39),
\beq
\rho =3p+\frac{(T_{\triangle V}\triangle )M^3k_BT}{2\pi ^3}\sum _{n=1}
^\infty \frac{(\pm 1)^{n+1}
}{n}e^{\frac{n\mu ^{'}}{k_BT}}K_1\left(\frac{nM}{k_BT}\right).
\eeq
The three relations (38),(40),(42) coincide with the corresponding
relations of an on-mass-shell relativistic kinetic theory (see Appendix
A), except for the additional factor $(T_{\triangle V}\triangle )/\pi .$
Therefore, (both the on-shell and the off-shell in the limit $m^2\cong
M^2)$ approaches give the same results, if one takes $T_{\triangle V}
\triangle =\pi .$ Since $2\triangle =\delta m$ is the width of the mass
deviation from its sharp value $m^2\cong M^2,$ the latter relation can be
rewritten as
\beq
T_{\triangle V}\delta m=2\pi .
\eeq

The presence of the factor $(T_{\triangle V}\triangle )/\pi $ in the
expressions (38),(40) and (42) shows that, as $\delta m\rightarrow 0,$
the corresponding quantities would go to zero, unless $T_{\triangle V}
\rightarrow \infty .$ Preservation of finite values of $p,\;\rho ,
\;N_0$ in this singular limit implies that the presence of events in the
ensemble with very sharp mass requires the extension of the set of events
over a very wide range of $\tau ,$ and therefore, $t,$ since they are
related through the equation of motion \cite{HP} $$\frac{dt}{d\tau }=
\frac{E}{M};$$ hence
\beq
\delta t=\langle dt\rangle =\langle d\tau \rangle \frac{\langle E\rangle
}{M}=T_{\triangle V}\frac{\sum _{n=1}^\infty \{(\pm 1)^{n+1}e^{\frac{n
\mu ^{'}}{k_BT}}K_2(\frac{nM}{k_BT})/n\}}{\sum _{n=1}^\infty
\{(\pm 1)^{n+1}e^{\frac{n\mu ^{'}}{k_BT}}K_1(\frac{nM}{k_BT})/n\}},
\eeq
where we used the relation $\langle d\tau \rangle =T_{\triangle V}$ and
the formula (16). In the Galilean limit $c\rightarrow \infty $ the
argument of the $K$-functions goes to infinity\footnote{This argument is
actually equal to $\frac{nMc^2}{k_BT}.$}, so that one obtains, using the
asymptotic formula (33) for $z\rightarrow \infty ,$
\beq
\langle dt\rangle =\langle d\tau \rangle ,
\eeq
in agreement with [6],[7].

If $\delta m$ is not so small, the range of $\tau ,$ and therefore, $t,$
need not be very large. In this case, for each value of the historical
time parameter $\tau ,$ the set of events making up the ensemble may be
concentrated in a range of $t$ which is not too large. The entire
ensemble then moves in phase space with the development of $\tau ,$ and
its stationary form corresponds to equilibrium. For $\delta m\rightarrow
0$ the ensemble fills a longer tube in phase space, and its displacement
with $\tau $ is not as important. It is in this limit alone that a direct
comparison with the usual (on-mass-shell) relativistic ensemble can be
made, since in this case the system possesses a stationarity in
space-time but
not a non-trivial evolution in $\tau ,$ as was remarked in ref. [4].

\section{Limiting cases}
Now we wish to consider the expressions for $N_0,\;p$ and $\rho $ in the
limiting cases $M\rightarrow 0,$ $T\rightarrow \infty ,$ and
$T\rightarrow 0,$ and to compare them with the corresponding formulas
of an on-mass-shell relativistic kinetic theory.

(i) $M\rightarrow 0.$

We remark that the dynamical evolution of the system contains the
quantity $\frac{p^\mu p_\mu }{2M}$ [15].  In the limit $M\rightarrow
\infty ,$ this quantity is undefined; however, if we first go to the
mass-shell limit, for which there is no non-trivial evolution in $\tau ,$
then this quantity becomes numerical-valued $(-M/2)$ and the limit $M
\rightarrow 0$ is well-defined. We therefore consider, for this limit,
the {\it on-shell} results (38),(40),(42),
obtained from the distribution function (35).

Using the asymptotic formula (33) for $z\rightarrow 0,$ one obtains
the relations
\beq
N_0=\pm \frac{(T_{\triangle V}\triangle )(k_BT)^3}{\pi ^3}Li_3(\pm e^{
\frac{\mu ^{'}}{k_BT}}),
\eeq
\beq
p=\pm \frac{(T_{\triangle V}\triangle )(k_BT)^4}{\pi ^3}Li_4(\pm e^{
\frac{\mu ^{'}}{k_BT}}),
\eeq
\beq
\rho =\pm 3\frac{(T_{\triangle V}\triangle )(k_BT)^4}{\pi ^3}Li_4(\pm e^{
\frac{\mu ^{'}}{k_BT}}),
\eeq
(here $Li_{\nu }(z)\equiv \sum _{k=1}^\infty \frac{z^k}{k^\nu }$ is the
so-called $polylogarithm$ \cite{Pru}), which coincide with the
corresponding relations of an on-shell theory, if $T_{\triangle V}
\triangle =\pi $ (see Appendix B).

(ii) $T\rightarrow \infty .$

This limit is mathematically equivalent to the limit $M\rightarrow 0;$
physically it means that the particles become distinguishable and possess
the relativistic distribution found by Synge \cite{Syn} .
Thus, one obtains the same expressions without summation on $k:$
\beq
N_0=\frac{(T_{\triangle V}\triangle )(k_BT)^3}{\pi ^3}e^{\frac{\mu ^{'}}
{k_BT}},
\eeq
\beq
p=\frac{(T_{\triangle V}\triangle )(k_BT)^4}{\pi ^3}e^{\frac{\mu ^{'}}
{k_BT}},
\eeq
\beq
\rho =3\frac{(T_{\triangle V}\triangle )(k_BT)^4}{\pi ^3}e^{\frac{\mu
^{'}}{k_BT}},
\eeq
in agreement with the textbook results, if $T_{\triangle V}\triangle =\pi
$ \cite{KT}.

(iii) $T\rightarrow 0.$

We study here the case in which the theory is {\it on-mass-shell.} If, in
the limit $T\rightarrow 0$ we do not insist on the mass-shell constraint,
then the mass distribution remains essentially relativistic. This case
was treated for distinguishable events in ref. [7]. The indistinguishable
case is treated in ref. [13].

Using the asymptotic formula (33) for $z\rightarrow \infty ,$ one obtains
\beq
N_0=\pm \frac{T_{\triangle V}\triangle }{\pi }\left(\frac{Mk_BT}{2\pi }
\right)^{3/2}Li_{3/2}(\pm e^{\frac{\mu ^{'}-M}{k_BT}}),
\eeq
\beq
p=\pm \frac{T_{\triangle V}\triangle }{\pi }\left(\frac{Mk_BT}{2\pi }
\right)^{3/2}k_BTLi_{5/2}(\pm e^{\frac{\mu ^{'}-M}{k_BT}})<<\rho ,
\eeq
\beq
\rho =\pm \frac{T_{\triangle V}\triangle }{\pi }\left(\frac{Mk_BT}{2\pi }
\right)^{3/2}\left[MLi_{3/2}(\pm e^{\frac{\mu ^{'}-M}{k_BT}})+3k_BT
Li_{5/2}(\pm e^{\frac{\mu ^{'}-M}{k_BT}})\right]=MN_0+3p,
\eeq
the formulas of a $nonrelativistic$ statistical mechanics of
indistinguishable particles\footnote{Indeed, using the formula (B.1) of
Appendix B, one has $(\epsilon \equiv \frac{{\bf p}^2}{2m})$ $$N_0=
\int \frac{d^3{\bf p}}{(2\pi )^3}\frac{1}{e^{(\frac{{\bf p}^2}{2m}-\mu )
/k_BT}\mp 1}=\frac{m^{3/2}}{2^{1/2}\pi ^2}\int \frac{\epsilon ^{1/2}d
\epsilon }{e^{(\epsilon -\mu )/k_BT}\mp 1}=\pm \left(\frac{mk_BT}{2\pi }
\right)^{3/2}Li_{3/2}(\pm e^{\frac{\mu }{k_BT}}),$$ the normalization
relation of a nonrelativistic statistical mechanics of indistinguishable
particles, etc. \cite{Hu}.}$^{,}$\footnote{See the corresponding
formulas of a nonrelativistic
statistical mechanics of distinguishable particles in ref. [18], p.62,
formulas (3.55).}, if one takes $T_{\triangle V}\triangle =\pi
$ and $\mu ^{'}-M\cong \mu -M=\mu _{nr},$ the nonrelativistic chemical
potential \cite{HW}.

As was remarked in [7], the low-temperature limit does $not$ necessarily
coincide with the nonrelativistic limit of a theory (for example, very
long wave length radiation in Maxwell's relativistic theory does not
coincide with the static Coulomb limit; it is still radiation). We see,
that the procedure of taking the limit $T\rightarrow 0$ on {\it mass
shell} leads to a deformation of the symmetry group from Poincar\'{e} to
Galilean invariance, which means that this procedure actually
coincides with the {\it Galilean limit} $c\rightarrow \infty $ \cite{AHW}
(formally equivalent to the limit $T
\rightarrow 0,$ since in both cases the argument of the $K$-functions in
the formulas (38),(40) and (42), $z\equiv \frac{nMc^2}{k_BT},$ goes to
infinity). We shall return to a more detailed study of this point in a
future research.

\section{Nonrelativistic distribution}
In conclusion we shall show that in the Galilean limit the variable
$E-m=\frac{{\bf p}^2}{2M}$ has the usual nonrelativistic
distribution for indistinguishable particles.

We start with the normalization relation for the sharp-mass form $m^2
\cong M^2$ of initial equilibrium relativistic distribution (4),
\beq
n(q)=C(q)\int d^4p\frac{e^{-AM^2-Am_c^2+2App_c+B}}{1\mp
e^{-AM^2-Am_c^2+2App_c+B}}.
\eeq
This integral written in the local rest frame takes the form
\beq
\int dEd^3{\bf p}\sum _{n=1}^\infty (\pm 1)^{n+1}e^{-2nAm_cE}e^{-n(AM^2+
Am_c^2-B)}\equiv \int dEd^3{\bf p}\sum _{n=1}^\infty (\pm 1)^{n+1}e^{-
\frac{nE}{k_BT}}e^{\frac{n\mu ^{'}}{k_BT}},
\eeq
or, in view of (2),(3) and the relation $-\triangle \leq \eta \leq
\triangle ,$
\beq
\sum _{n=1}^\infty \int _{M-\triangle }^{M+\triangle }
dm\;e^{-\frac{nm}{k_BT}}\int d^3{\bf p}e^{-\frac{n}{k_BT}
(\frac{{\bf p}^2}{2M}-\mu ^{'})}.
\eeq
Integration on $m$ gives
\beq
\int _{M-\triangle }^{M+\triangle }dm\;e^{-\frac{nm}{k_BT}}=
2\triangle e^{-\frac{nM}{k_BT}};
\eeq
therefore
\beq
n(q)=2\triangle
C(q)\int d^3{\bf p}\sum _{n=1}^\infty (\pm 1)^{n+1}e^{-\frac{n}{k_BT}(
\frac{{\bf p}^2}{2M}-\mu ^{'}+M)}\equiv 2\triangle C(q)\int \!d^3{\bf p}
\;\frac{1}{e^{(\frac{{\bf p}^2}{2M}-\tilde{\mu })/k_BT}\mp 1},
\eeq
which is the usual (normalized) nonrelativistic Bose-Einstein/Fermi-Dirac
distribution (we call $e\equiv \frac{{\bf p}^2}{2M})
$
\beq
f(e)=\tilde{D}\frac{e^{1/2}}{\exp (\frac{e-\tilde{\mu }}{k_BT})\mp 1},
\eeq
\beq
\tilde{D}^{-1}=\Gamma \left(\frac{3}{2}\right)(k_BT)^{3/2}\sum _{n=1}^
\infty \frac{(\pm 1)^{n+1}}{n^{3/2}}e^{\frac{n\tilde{\mu }}{k_BT}}.
\eeq
In this formulas
\beq
\tilde{\mu }=\mu ^{'}-M=\mu -M\left(1+\frac{\mu _K}{2N}\right)
\eeq
is the ``reduced'' nonrelativistic chemical potential approaching, as
$N\rightarrow \infty $ or $T\rightarrow 0,$ the value $(\mu -M)$
[1],[19].

\section{Concluding remarks}
We have considered the limiting
mass-shell form of equilibrium relativistic
distribution for indistinguishable events, studied previously in [1], and
its Galilean limit. We have shown that the mass-shell forms of the basic
thermodynamic quantities coincide with the corresponding expressions of a
usual on-shell relativistic statistical mechanics, providing the
important relation (43) for a width of the mass deviation over its sharp
value. This relation implies that such a mass deviation in the system of
events can be infinitely small. In this case the system of relativistic
(mass shell) events becomes a stationary distribution in space-time,
representing in this way the ensemble of the
relativistic on-shell particles.

In ensembles, in which the appreciable mass fluctuations can occur, the
relativistic distributions (the equilibrium solutions of the generalized
Boltzmann equation) represent the corresponding equilibrium relativistic
distributions of mass, considered previously in refs. [1],[4], for
identical and nonidentical systems, respectively.

We have seen that the characteristic limiting cases give results in
agreement with the corresponding limits of the usual theory; the
low-temperature limit on mass shell in fact corresponds to the
Galilean limit $c\rightarrow \infty ,$ in which the statistical mechanics
of indistinguishable events goes over to the usual
nonrelativistic statistical mechanics of indistinguishable particles.

For ensembles with mass fluctuations the framework of a
manifestly covariant relativistic statistical mechanics provides
corrections to results obtained within the usual on-shell theory.
Physical consequences of the mass fluctuations for relativistic
systems are considered in refs. [13],\cite{cond},\cite{prep}, taking
account $anti$-$events,$ i.e., the events having the opposite sign of
the mass potential $\mu _K.$ In ref. [13] an adiabatic equation of
state, $p\propto N_0^{6/5},$ is obtained for the system of degenerate
off-shell fermions and possible implications in astrophysics are
discussed. In \cite{cond} statistical mechanics of the bosonic
event--anti-event system is considered. For such a system, at some
critical temperature a special type of
Bose-Einstein condensation sets in, which provides
the events making up the ensemble a definite mass and represents, in this
way, a phase transition to the usual on-shell sector. In ref. \cite{prep}
possible applications of the off-shell theory in hadronic physics are
discussed. The equation of state $p\propto T^6,$ obtained in \cite{prep},
corresponds to the ``realistic'' equation of state, proposed by
Shuryak \cite{Shu} for hot hadronic matter, which is in good agreement
with experiment for the temperature range 0.2--1.0 GeV.

\newpage
\appendix
\section*{Appendix A}
\setcounter{equation}{0}
\renewcommand{\theequation}{A.\arabic{equation}}
Consider the following relation, which represents the normalization
condition for the relativistic distribution function $$f_0(|{\bf p}|,m)=
\frac{1}{e^{(\sqrt{{\bf p}^2+m^2}-\mu )/k_BT}\mp 1}$$ ($\epsilon \equiv
\sqrt{{\bf p}^2+m^2}):$
\beq
N_0=\int \frac{d^3{\bf p}}{(2\pi )^3}\frac{1}{e^{(\sqrt{{\bf p}^2+m^2}-
\mu )/k_BT}\mp 1}=\frac{1}{2\pi ^2}\int _m^\infty \!d\epsilon \;\frac{e^{
-(\epsilon -\mu )/k_BT}\epsilon \sqrt{\epsilon ^2-m^2}}{1\mp e^
{-(\epsilon -\mu )/k_BT}}.
\eeq
Expanding the denominator into power series and changing variables $t=
\epsilon /m,$ one obtains
\beq
N_0=\frac{m^3}{2\pi ^2}\sum _{n=1}^\infty (\pm 1)^{n+1}e^{\frac{n\mu }{k_
BT}}\int _1^\infty \!dt\;t\sqrt{t^2-1}e^{-\frac{nm}{k_BT}t}.
\eeq
Since \cite{GrRy4}
\beq
\int _1^\infty \!dt\;\sqrt{t^2-1}e^{-\frac{nm}{k_BT}t}=\frac{k_BT}{nm}K_1
\left(\frac{nm}{k_BT}\right),
\eeq
one uses parametric differentiation with respect to $(nm/k_BT)$ and Eq.
(10) of the main text to obtain
\beq
\int _1^\infty \!dt\;t\sqrt{t^2-1}e^{-\frac{nm}{k_BT}t}=-\frac{\partial }
{\partial (\frac{nm}{k_BT})}\int _1^\infty \!dt\;\sqrt{t^2-1}
e^{-\frac{nm}{k_BT}t}=\frac{k_BT}{nm}K_2\left(\frac{nm}{k_BT}\right).
\eeq
Thus,
\beq
N_0=\frac{m^2k_BT}{2\pi ^2}\sum _{n=1}^\infty \frac{(\pm 1)^{n+1}}{n}e^{
\frac{n\mu }{k_BT}}K_2\left(\frac{nm}{k_BT}\right).
\eeq
In the same way one can obtain \cite{KT1}
\beq
p=\!\!\int \!\!\frac{d^3{\bf p}}{(2\pi )^3}\frac{{\bf p}^2}{3\sqrt{{\bf
p}^2+m^2}}\frac{1}{e^{(\sqrt{{\bf p}^2+m^2}-\mu )/k_BT}\mp 1}=\frac{m^2
(k_BT)^2}{2\pi ^2}\!\!\sum _{n=1}^\infty \!\frac{(\pm 1)^{n+1}}{n^2}
e^{\frac{n\mu }{k_BT}}K_2\!\left(\frac{nm}{k_BT}\right)\!,
\eeq
\beq
\rho =\int \frac{d^3{\bf p}}{(2\pi )^3}\frac{\sqrt{{\bf p}^2+m^2}}
{e^{(\sqrt{{\bf p}^2+m^2}-\mu )/k_BT}\mp 1}=3p+
\frac{m^3k_BT}{2\pi ^2}\sum _{n=1}^\infty \frac{(\pm 1)^{n+1}}{n}
e^{\frac{n\mu }{k_BT}}K_1\left(\frac{nm}{k_BT}\right).
\eeq

\newpage
\appendix
\section*{Appendix B}
\setcounter{equation}{0}
\renewcommand{\theequation}{B.\arabic{equation}}
First consider
\bqry
\int _0^\infty \frac{a^\nu da}{e^{a-b}\mp 1} & = & \int _0^\infty \!
da\;a^\nu \frac{e^{-a+b}}{1\mp e^{-a+b}}=\sum _{k=1}^\infty
(\pm 1)^{k+1}e^{kb}\int _0^\infty \!da\;a^\nu e^{-ka} \NL
  & = & \Gamma (\nu +1)\sum _{k=1}^\infty \frac{(\pm 1)^{k+1}}{k^{\nu +1}
}e^{kb}=\pm \Gamma (\nu +1)Li_{\nu +1}(\pm e^b),
\eqry
where $Li_n(z)\equiv \sum _{k=1}^\infty \frac{z^k}{k^n}$ is the
polylogarithm [16].
Then
\beq
N_0=\int \frac{d^3{\bf p}}{(2\pi )^3}\frac{1}{e^{(|{\bf p}|-\mu )/k_BT}
\mp 1}=\frac{4\pi }{(2\pi )^3}\int \frac{p^2dp}{e^{(p-\mu )/k_BT}\mp 1}=
\pm \frac{(k_BT)^3}{\pi ^2}Li_3(\pm e^{\frac{\mu }{k_BT}}),
\eeq
\beq
p=\int \frac{d^3{\bf p}}{(2\pi )^3}\frac{{\bf p}^2}{3|{\bf p}|}
\frac{1}{e^{(|{\bf p}|-\mu )/k_BT}
\mp 1}=\frac{4\pi }{3(2\pi )^3}\int \frac{p^3dp}{e^{(p-\mu )/k_BT}\mp 1}=
\pm \frac{(k_BT)^4}{\pi ^2}Li_4(\pm e^{\frac{\mu }{k_BT}}),
\eeq
\beq
\rho =\int \frac{d^3{\bf p}}{(2\pi )^3}\frac{|{\bf p}|}{e^{(|{\bf p}|-\mu
)/k_BT}\mp 1}=\frac{4\pi }{(2\pi )^3}\int \frac{p^3dp}{e^{(p-\mu )/k_BT}
\mp 1}=\pm 3\frac{(k_BT)^4}{\pi ^2}Li_4(\pm e^{\frac{\mu }{k_BT}}).
\eeq

\end{document}